# Validation of Enhanced Emotion Enabled Cognitive Agent Using Virtual Overlay Multi-Agent System Approach

Faisal Riaz[1], Muaz A. Niazi[2]

[1]Dept. Of Computer Sciences & IT-Mirpur University of Science & Technology, AJ&K, Pakistan
[1]Dept. Of Computing and Technology-Iqra University, Islamabad, Pakistan
[2]Dept. Of Computer Sciences-COMSATS, Islamabad, Pakistan
Email: *fazi_ajku@yahoo.com, muaz.niazi@gmail.com

**Abstract**

Making roads safer by avoiding road collisions is one of the main reasons for inventing Autonomous vehicles (AVs). In this context, designing agent-based collision avoidance components of AVs which truly represent human cognition and emotions look is a more feasible approach as agents can replace human drivers. However, to the best of our knowledge, very few human emotion and cognition-inspired agent-based studies have previously been conducted in this domain. Furthermore, these agent-based solutions have not been validated using any key validation technique. Keeping in view this lack of validation practices, we have selected state-of-the-art Emotion Enabled Cognitive Agent (EEC_Agent), which was proposed to avoid lateral collisions between semi-AVs. The architecture of EEC_Agent has been revised using Exploratory Agent Based Modeling (EABM) level of the Cognitive Agent Based Computing (CABC) framework and real-time fear emotion generation mechanism using the Ortony, Clore & Collins (OCC) model has also been introduced. Then the proposed fear generation mechanism has been validated using the Validated Agent Based Modeling level of CABC framework using a Virtual Overlay MultiAgent System (VOMAS). Extensive simulation and practical experiments demonstrate that the Enhanced EEC_Agent exhibits the capability to feel different levels of fear, according to different traffic situations and also needs a smaller Stopping Sight Distance (SSD) and Overtaking Sight Distance (OSD) as compared to human drivers.

# 1. Introduction

Making roads safer by avoiding road collisions is one of the main reasons for inventing Autonomous vehicles (AVs) [1]. Vine et al. [2] have proposed collision free intersections using autonomous cars by employing the clear distance ahead approach. Rodrıguez- Seda et al. [3] have developed collision–free maneuvers in uncertain road environments for autonomous cars using Lyapunov-based analysis. Jimenez et al. [4] have made an autonomous car capable of avoiding road collisions from pedestrians and surrounding vehicles using laser-scanner sensor and detailed digital map. However, designing collision avoidance component of AVs inspired by agent based modeling, which truly represent human cognition and emotions look more feasible as they are replacing human drivers.

Researchers have proposed many agent based collision avoidance systems. Reichardt [5] has presented an emotional agent inspired driver's assistant model which simulates the emotional influence on the human driver's behavior. The main purpose of this model is to build a framework for learning algorithms, which help in building adaptive driver assistance system. An Emotion Enabled Cognitive Agent (EEC_Agent) inspired lateral collision avoidance scheme between AVs has been proposed by Riaz et al. [6]. A detailed analysis of Cyberphysical systems for collision avoidance has also been proposed by Riaz and Niazi in [7]. An in-vehicle virtual agent based driver assistance system has been proposed by Joo and Lee-Won [8], which help the female drivers to improve their performance during risky situations. A human behavior inspired agent has been proposed by Waizman et al. [9], which help the AVs to avoid the collisions from the pedestrians at black spots. An agent based Driver Assistance System (DAS) has been proposed by Tiengo et al. [10] for the rollover prevention in the heavy duty vehicles. In another work, a rule based cognitive Agent inspired intersection collision avoidance system has been proposed by Lu et al. [11]. However, the problem with these agent based solutions is that, the claimed functionalities of these agents based systems have not been validated at any level.

In existing literature, different validation techniques for agent based systems have been proposed by the researchers. These include work on modeling the internet of things such as [12-14]. For complex-network based models, a validation methodology has also been proposed by Batool and Niazi in [15]. Agents have also been demonstrated to be useful in the domain of multi-agent foraging [16]. Fagiolo et al. [17] have proposed empirical validation technique to validate the agent-based systems. In another research work, philosophical truth theories based validation scheme has been proposed by Schmid [18]. Barreteau et al. [19] have proposed iterative participatory approach known as companion modeling for the validation of agent based systems. Makowsky [20] has proposed agent based simulation itself as a validation technique to validate the functionality of proposed agent system. Niazi et. al [21] have proposed a novel concept of agent based validation using Virtual Overlay Multi-Agent System (VOMAS) under the framework of Cognitive Agent Based Computing (CABC) framework [22]. However, any of these above mentioned validation techniques have been not applied to validate the such cognitive and emotional agents , which have been tailored specialy for autonomous vehicles to enhance their collision avoidance capabilities.

Contribution: The main contribution of this research work is to validate the existing state-of-the-art EEC_Agent proposed by Riaz et al. [6]. However, some secondary contributions have been made as well, which were necessary to perform its validation. The details of the contribution are given as under.

1) The enhanced version of EEC_Agent known as Enhanced Emotion Enabled Cognitive Agent (EEEC_Agent) has been proposed by introducing proper emotion generation mechanism using Ortony, Clore & Collins OCC model [23], as it lacks in simple EEC_Agent [6]. For this purpose the Exploratory Agent Based Modeling (EABM) level of CABC framework has been employed.

2) The validated Agent Based Modeling level of CABC framework has been utilized to validate the EEEC_Agent functionalities. To validate the EEEC_Agent Virtual Overlay Multi-Agent System (VOMAS) approach has been employed for comparing and validating the performance of EEEC_Agent with that of a human driver during the rear end collision situation. For comparison and validation of EEEC_Agent, Stopping Sight Distance (SSD) defined by the American Association of State Highway and Transportation Officials (AASHTO) [24] and Overtaking Sight Distance (OSD) defined by Indian Road Congress (IRC) [25] are used.

The extensive experiments prove that EEEC_Agent enabled AVs can avoid rear end and lane changing collisions with smaller SSDs and OSDs respectively as compared to the human-driven vehicles. Hence, then validated and truly emotional, cognitive agent-based collision avoidance solution for the autonomous vehicles is revealed.

The rest of the paper has been organized as follows. Section 2 presents the background. Section 3 discusses EABM based design of proposed EEEC_Agent. SimConnect design of the proposed EEEC_Agent simulation has been presented in section 4. Section 5 presents the validation model. Section 6 presents experimental setup. Section 7 gives results and discussion along with the details of experiments. Practical validation of the proposed fear generation mechanism of EEEC_Agent has been performed in section 8 using prototype AV. The paper concludes in Section 9.

## 2. Background

This background section gives the preliminary information about the key terms utilized in this research work. First of all, the role of emotions in human life has been elucidated along with the OCC model of emotions. Subsequently, the terms sight distances, Cognitive agent based computing, SimConnector and VOMAS agent have been discussed.

### a. Emotions

The term 'emotion' has been used to refer mental and physical course of action that includes aspects of subjective experience, evaluation & appraisal, motivation and body responses such as arousal and facial expression [26]. According to Aristotle, emotion is defined as a hat that leads one's state to become so transformed that his judgment is affected, which is accompanied by pleasure or pain [26] .There are several emotion models, which are proposed by different researchers. However, we are interested in the OCC model [23].The reason for choosing OCC is the primary interest of its authors in the role of cognition which is used to generate emotion. According to OCC model, the emotions may be generated due to three major aspects of the world or changes in the world, namely events, agents, or objects. When humans focus on events they are interested in their consequences, whereas they focus on agents and actions, they are interested in their actions. In our work, we are interested in emotions generated due to the reactions to the expected events. These types of emotions are also known as prospect based emotions.

### b. Cognitive Agent-Based Computing (CABC)

Agent-based modeling (ABM) and complex networks (CN) are two popular modeling tools for understanding Complex Adaptive System (CAS). In 2011, a unified framework named Cognitive Agent-based Computing (CABC) combining these two modeling paradigms was proposed by Muaz et al. [22] for the better understanding of CAS. Agent based modeling (ABM) and complex networks (CN) are two popular modeling tools for understanding

Complex adaptive system (CAS). The CABC helps the cross-disciplinary researchers to develop the understanding of their area related CAS using different types of models. It provides guidelines to the multidisciplinary researchers regarding how they can develop computational models of CAS even they belong to social science, life science or computer science. The unified framework provides four understanding and development levels of CAS along with related case studies.

**Complex Network Modeling.** The first level of framework which is useful in modeling Complex systems is complex network modeling. When the interaction data between network nodes is available then using complex network modeling can be useful. This level helps the researchers in building the complex network models along with the network classification. Further this level helps in extracting the useful information from the network by determining the global and local quantitative measures related to this network. Other statistical and more traditional mathematical models have not such capability to provide details of emergent behavior and patterns of complex networks, which can be achieved by complex network modeling.

**Exploratory Agent Based Modeling.** The second level of framework is Exploratory Agent Based Modeling (EABM). When the researchers are interested in extending existing ideas related to the agent based modeling belonging to the other fields, the EABM is a useful guideline paradigm in this regard. Using EABM, researchers can build experimental or proof of concepts, which help in defining the further scope and feasibility of the future research. Using EABM researchers still failed to solve some important problems.

**DREAM.** Using DREAM level, quantitative comparison of different models can be performed without execution of simulation experiments. The processes of reverse engineering and replication of model can be easily done using DREAM. Hence the ABM can be examined visually other than textual description. The visual examination of ABM is better in sense that model can be analyzed abstractly without visiting its source code. Hence using visual approach the comparison between domains across models can be done easily along with teaching cas models. Another ultimate benefit of DREAM is translation of different sub-models from visual model to pseudo code specification model to an agent-based model. Thirdly the proposed methodology should allow for a translation from these different sub-models such as from a visual (Complex network-based) model to pseudo code specification model to an agent-based model.

**Validated Agent Based Modeling.** Verification and validation are the techniques used for the evaluation of a product or system that it meets its objectives, requirements and specification. These approaches are used together but they are different from each other. The validated agent-based modeling level of the proposed framework is concerned with developing verified and validated agent-based models. This level allows performing in-simulation verification and validation of the agent-based models using a Virtual Overlay Multi-agent System (VOMAS). To solve this problem Muaz et.al proposed a novel concept of Virtual Overlay Multi-Agent System (VOMAS) in [28]. One of core benefits of VOMAS is that all kinds of agent based models can be validated by using it. The basic idea of VOMAS is performing validation by making an overlay on the top of agent based simulation without taking an active part in the simulation. Using VOMAS, agent based simulations can be validated both spatially and non-spatially.

In this research paper we have utilized two levels of CABC framework. The EABM, which help us in exploring the role of the OCC model in enhancing the emotion generation capabilities of EEEC_Agent and Validated agent based modeling level, which provides us guidelines to validate the functionality of EEEC_Agent.

### c. Sight Distances

Sight distances are the different types of safe distances between the vehicles which should be maintained to avoid the collisions. Sight distances are further subdivided in stopping sight distance and overtaking sight distance.

**Stopping Sight Distance (SSD).** The minimum sight distance at any point which enables the driver to stop the vehicle safely without the collision is known as stopping sight distance. It is one of the very basic measures in traffic engineering defined by the (AASHTO) [27]. SSD basically comprises of two distances. The first one is the distance that vehicle travels during the reaction time of the driver and the second one is the distance that vehicle takes to stop after applying the brakes.

**Overtaking Sight Distance (OSD).** The Overtaking Sight Distance (OSD) is the distance opens in front of the driver of a vehicle while trying to overtake against opposite direction. IRC has given a method to compute OSD in[28].

### d. VOMAS

Agent-based models are getting popular due to their easy application to the different fields of life, such as modeling human muscle development, heat diffusion, Turing machine, flocking and different social science mechanisms, etc. In spite of such popularity of agent-based model, validation of these models is still a challenging task. During the modeling of agent-based models, it is important to ensure that the model is working correctly, i.e. verification and the model is giving required outputs i.e. validation. To solve this problem, Muaz et al. [29] proposed a novel concept of VOMAS. Using VOMAS, agent-based simulations can be validated both spatially and non-spatially. In our work for the validation of proposed agent-based model of EEC_Agent, we are using multi-invariant based validation method, which lies in non-spatial validation category.

### e. SimConnector

As the natural disaster events occur rarely so testing the performance of disaster alert systems is a challenging task. To overcome this problem, a novel approach was proposed by Muaz et.al in [30] by developing and testing real-time disaster early warning and alerting system by combining two different software environments. In [30], the agent-based simulation was used to generate a rare forest fire event and a web-based alert decision support system for generating warnings. For a detailed study of SimConnector, the interested readers are referred to [30].

## 3. Improvement in the Existing Architecture of EEC_Agent Using EABM

In the introduction section, it has already been discussed that in the current work we are going to improve the existing architecture of EEC_Agent [6]. The main drawback of EEC_Agent is that it is claiming utilization of emotions in making collision avoidance decisions, but no proper emotion generation mechanism has been proposed, which helps the cognitive agent to feel emotion according to the changing in the dynamic environment. Hence to overcome this issue, we have redesigned the architecture of EEC_Agent using EABM and explored the role of OCC model in the fear generation of EEC_Agent. In existing literature OCC model has not been utilized to model the collision avoidance problems of AVs using an agent approach. To our best knowledge, it is the first research work, which is going to explore the role of OCC model in this context. The architecture and functionality of the EEEC_Agent are discussed as follows.

### 3.1. Architectural Features and Functionality

The architecture of the proposed EEEC agent has been shown in figure 1. For the sake of less complexity only two main neurons are considered i.e. Hypothalamus neuron (NH) and Amygdala neuron (NA) because of their direct relation regarding the efficient processing in emergency situations. In addition to generate the notion of fear in an EEC agent, the OCC Model Based Fear Generation Module has been introduced [15].

It can be seen from the figure 1 that the proposed architecture consists of five main modules: Sensory module, Artificial Thalamus module, OCC Model Based Fear Generation Module, and Motor module. To begin with, step 1, the Sensory module keeps track of the distance between neighboring AVs on a road segment. The Sensory module of the EEC_Agent converts the stimulus information into electrical signals. The Artificial Thalamus module, similar to the hypothalamus module in the human brain, receives these sensed signals in step 2 for further processing. The signals received by Thalamus module have different frequencies that reflect on the prevailing inter-vehicular distance. In step 3, these signals are then checked against the maximum allowed threshold after computing likelihood, desirability and Ig variables using OCC Model Based Fear Generation Module. In step 4 and 5, the Artificial Amygdala module also computes the fear intensity level in coordination with Artificial Thalamus and OCC Model Based Fear Generation Modules. These different intensity levels are used as controlling interrupts, which are passed to the Motor Module, in step 6, and in turn it execute the suitable collision avoidance maneuver.

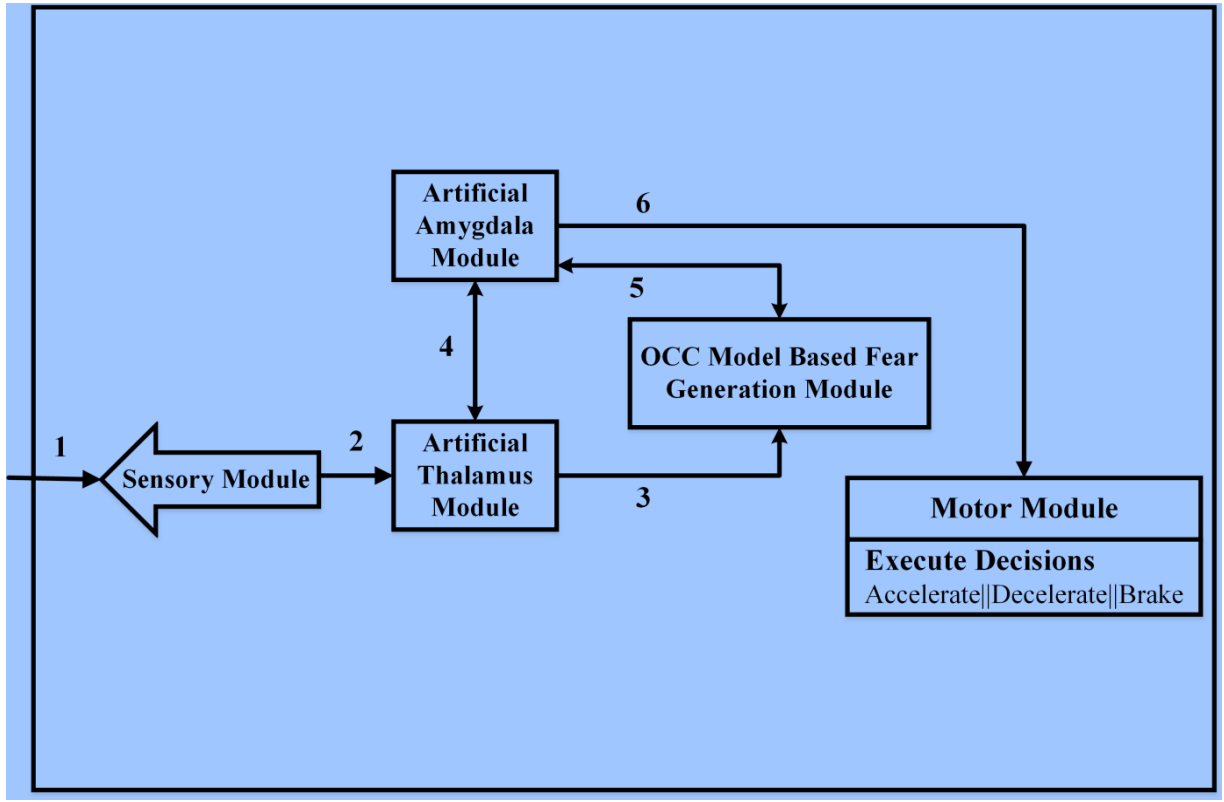

Figure. 1 Emotion Enabled Cognitive Agent Architecture

## 4. Proposed SimConnector Design For Implementation of EEEC_Agent

The SimConnector approach proposed in [30] consists of an agent-based simulation, which generates artificial disaster events and then provided this data to the web-based decision support system for generating warnings. However, in our case, agent-based simulation is used for validating the performance of prospect based emotion, i.e. fear inspired road collision avoidance system in real road accident scenarios. We first developed a fuzzy logic based simulator for finding the numeric values of fear related variables such as Desirability, Likelihood, and Ig. These numeric values are provided to the NetLogo platform based model of EEC_Agent for generating fear and testing its performance in rear-end collisions. Fig. 2 depicts our SimConnector design.

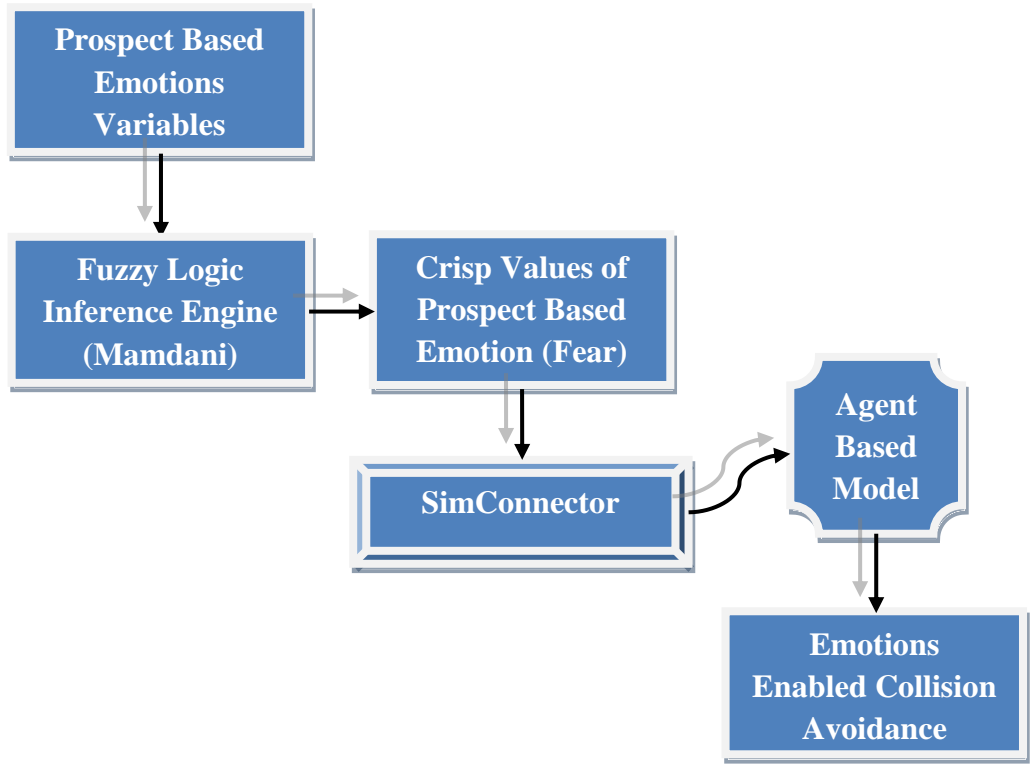

Figure. 2 SimConnector design for the implementation

## 5. EEEC_Agent Validation Model

As mentioned earlier, non-validated simulation and results are one of the drawbacks of previously proposed EEC_Agent [6]. To overcome this problem, we have proposed VOMAS agent-based simulation validation methodology. First, we have elucidated the VOMAS agent design, and then we have raised the validation question. In the last, three invariants have been defined which will act as filters to validate the agent-based simulation of EEEC_Agent.

**VOMAS Agent Design**- In our simulation model, we have designated AV as a VOMAS agent. The VOMAS agent computes the changes in AV's fear, according to the OCC model defined fear related variables and computes the required SSD and OSD distances required for efficient collision avoidance during rear end and overtaking scenarios respectively..

### 5.1 Validation Question And Invariants

Our validation question can thus be defined as. "How can we validate that the emotion generation mechanism of EEEC_Agent is working properly and EEEC_Agent installed AV avoid the rear end and overtaking collisions more eficiently than human drivers." To find out the answer of these validation questions Invariants methodology proposed by Niazi et al. [29] has been utilized and in this regard following four invariants have been defined. The first two invariants help in validating that the proposed fear generation mechanism of EEEC_Agent is working properly. The third and fourth invariants help in validating that EEEC_Agent installed AV requires smaller SSD and OSD during rear end collision avoidance and overtaking scenarios respectively as compared to the human drivers.

### a). Invariant1_TypeA

If the pre-condition that **"*Distance between the rear end of the first AV and the front end of the second AV is very small*"** is true, then the fear level exhibited by AV would result in a post-condition of **"*Intensity of a fear of a bullet autonomous vehicle is high or very high*"**.

### b). Invariant1_TypeB

If the pre-condition that **"*Distance between the rear end of the first AV and the front end of the second AV is decreasing*"** is true, then the variation in the distance of AV would result in a post-condition of **"*Intensity of a fear of a bullet autonomous vehicle is increasing accordingly*"**.

### c).Invariant2

If the pre-condition that **"*Bullet_Agent autonomous vehicle is successfully reacting to the rear end collision threat using EEEC_Agent short route reaction time*"** is true, then stopping sight distance required would result in a post-condition of **"*EEEC_Agent requires smaller SSD as compared to the SSD required by Human driver*"**.

### d). Invariant3

If the pre-condition that **"*Bullet_Agent autonomous vehicle is successfully reacting to the overtaking collision threat using EEEC_Agent short route reaction time*"** is true, then overtaking sight distance required would result in a post-condition of **"*EEEC_Agent requires smaller OSD as compared to the OSD of the human driver*"**.

## 6. Experiments

This section describes the experiments related to the quantitative computation of prospect-based emotion using fuzzy logic and the experiments related to the validation of EEC_Agent.

## 6.1 Experiment 1

To compute the fear, we have built a Mamdani fuzzy inference system, which uses the traceability algorithm defined in [13].

> If Prospect (v, e, t) and Undesirable (v, e, t) < 0
>
> Then set Fear-Potential (v, e, t) = ff [|**Desire (v, e, t)** |, **Likelihood (v, e, t), Ig (v, e, t)**]
>
> If *Fear-Potential* (v, e, t) > Fear-Threshold (v, t)
>
> Then set *Fear-Intensity* (v, e, t) = Fear-Potential (v, e, t) - Fear-Threshold

**Implementation Details Of Fuzzy Logic To Compute The Numeric Values of Fear Emotion**

For the sake of brevity, we are just giving here short details of the Mamdani fuzzy inference system for the computation of different intensities of fear. The details of Likelihood variable are given in the coming section along with Linguistic tokens and fuzzy rules.

**Sub-Experiment 1a**

The likelihood variable helps in computing the chances of an accident. Suppose that the AV is following a truck. If the distance between AV and the truck is higher or equal to the SSD than the likelihood of accidents will be low and vice versa. After a detailed analysis, it has been observed that distance and speed are two such factors which directly affect the likelihood of an accident. For example, if the speed is high and distance is low, then the likelihood of an accident will be high, and on the other hand, if the speed is low and distance is high then the likelihood of an accident will be low. The experiment 1a has been carried out to compute Likelihood variable using fuzzy inference editor as shown in Fig. 3. The two input variables Distance and Speed are defined on the left side and the Likelihood variable on the right side. The mathematical function of TRIMF has been utilized as a membership function for two input and one output variable. To compute the Likelihood variable, five linguistic tokens VLLH, LLH, MLH, HLH and VHLH have been defined which represent Very low likelihood, Low likelihood, Medium likelihood, High likelihood and Very High likelihood respectively.

Table I Linguistic tokens for Likelihood

| Linguistic tokens | Description |
|---|---|
| VHLH | Very High Likelihood |
| HLH | High Likelihood |
| MLH | Medium Likelihood |
| LLH | Low Likelihood |
| VLLH | Very Low Likelihood |

Twenty-five rules were defined to obtain the value of the variable likelihood, these rules are presented in table II.

Table II Fuzzy Rules defined for Likelihood

| If Distance is | And Speed is | Then Likelihood is |
|---|---|---|
| VHD | VHS | MLH |
| VHD | HS | LLH |
| VHD | MS | VLLH |
| VHD | LS | VLLH |
| VHD | VLS | VLLH |
| HD | VHS | HLH |
| HD | HS | MLH |
| HD | MS | VLLH |
| HD | LS | VLLH |
| HD | VLS | VLLH |
| MD | VHS | VHLH |
| MD | HS | VHLH |
| MD | MS | MLH |
| MD | LS | LLH |
| MD | VLS | VLLH |
| LD | VHS | VHLH |
| LD | HS | VHLH |
| LD | MS | HLH |
| LD | LS | MLH |
| LD | VLS | VLLH |
| V LD | VHS | VHLH |
| V LD | HS | VHLH |
| V LD | MS | VHLH |
| V LD | LS | HLH |
| V LD | VLS | MLH |

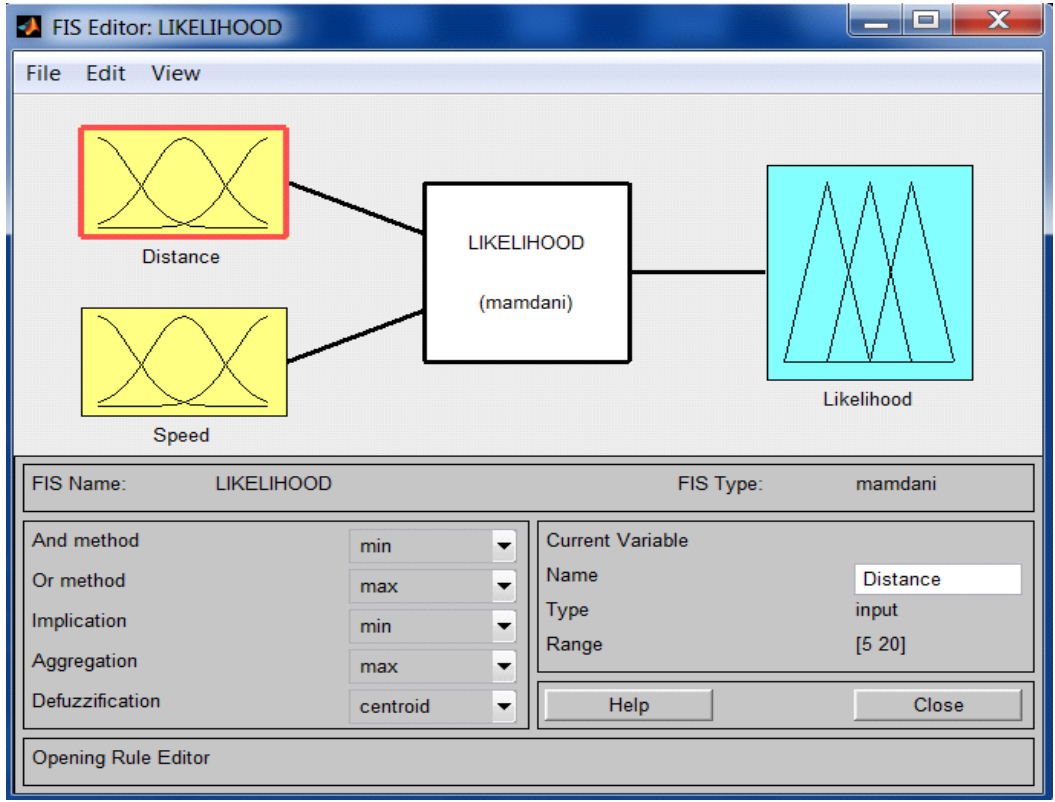

Figure 3 Main simulation screen for Likelihood computation

The remaining fear variables, i.e. Undesirability and Ig are computed on the same pattern.

**6.2 Experiment 2**

The purpose of the second type of experiment is to perform the validation of SimConnector based design of the EEEC_Agent using VOMAS agent. For this purpose, Netlogo 5.3 has been utilized which is a standard agent-based simulation environment. The NetLogo 5.1 environment consists of patches and turtles.

**6.2.1 Experimental Parameters**

To perform the simulation experiments empirically, three types of simulation parameter sets are defined. The first set consists of numeric values (Table 9, 10 and 11) of prospect based emotions (i.e. Fear) variables like the likelihood of accident event, the undesirability of accident event, and Ig. The second set of parameters consists of Stopping Sight Distance and overtaking sight distance described by eq. (1) and (2).

$$\text{SSD} = 1.47\text{Vt} + \frac{1.075V^2}{a} \qquad (1)$$

Where SSD = Stopping Sight Distance in feet, V = design speed in mph, t = brake reaction time in seconds and    a = deceleration rate, 11.2 ft/s2.

$$OSD = V_b\, t + 2s + V_b \sqrt{\frac{4s}{a}} \qquad (2)$$

Here Vb = velocity of overtaking Vehicle, t = Reaction time, S=space before and after overtaking and a = Maximum overtaking acceleration at different speeds.

The third miscellaneous set of parameters is.

-Number of Agents

-Number of VOMAS agents

-Testing Speed range (mph)

-EEEC_Agent Status

-Reaction Time

Here 0.4397 seconds and 3.8085 seconds are the reaction times for taking collision avoidance maneuver by EEEC_Agent and human driver respectively. These reaction times are taken in the guidelines of [6].

Fig.6 shows the experimental environment along with input and output parameters. The two AVs are taking part in this validation simulation. The bullet AV is acting as a VOMAS agent whereas the second one is leading AV which acts as a target agent. The left side of the simulation world contains input sliders for providing fuzzy logic based numeric values of prospect based emotion variables (Undesirability, Likelihood, Ig). It is important to recall here that these numeric values of prospect based emotions were computed through experiments a), b) and c), presented in section 5.1.1, using fuzzy logic and then provided to the agent based simulation using the  proposed SimConnector approach. The world size used in the simulation is (-25, -25) to (25, 25) and in this way the total number of patches in the world is 25. To map the real-world distance in feet on 25 patches, each patch is representing a value equal to 100 feet.

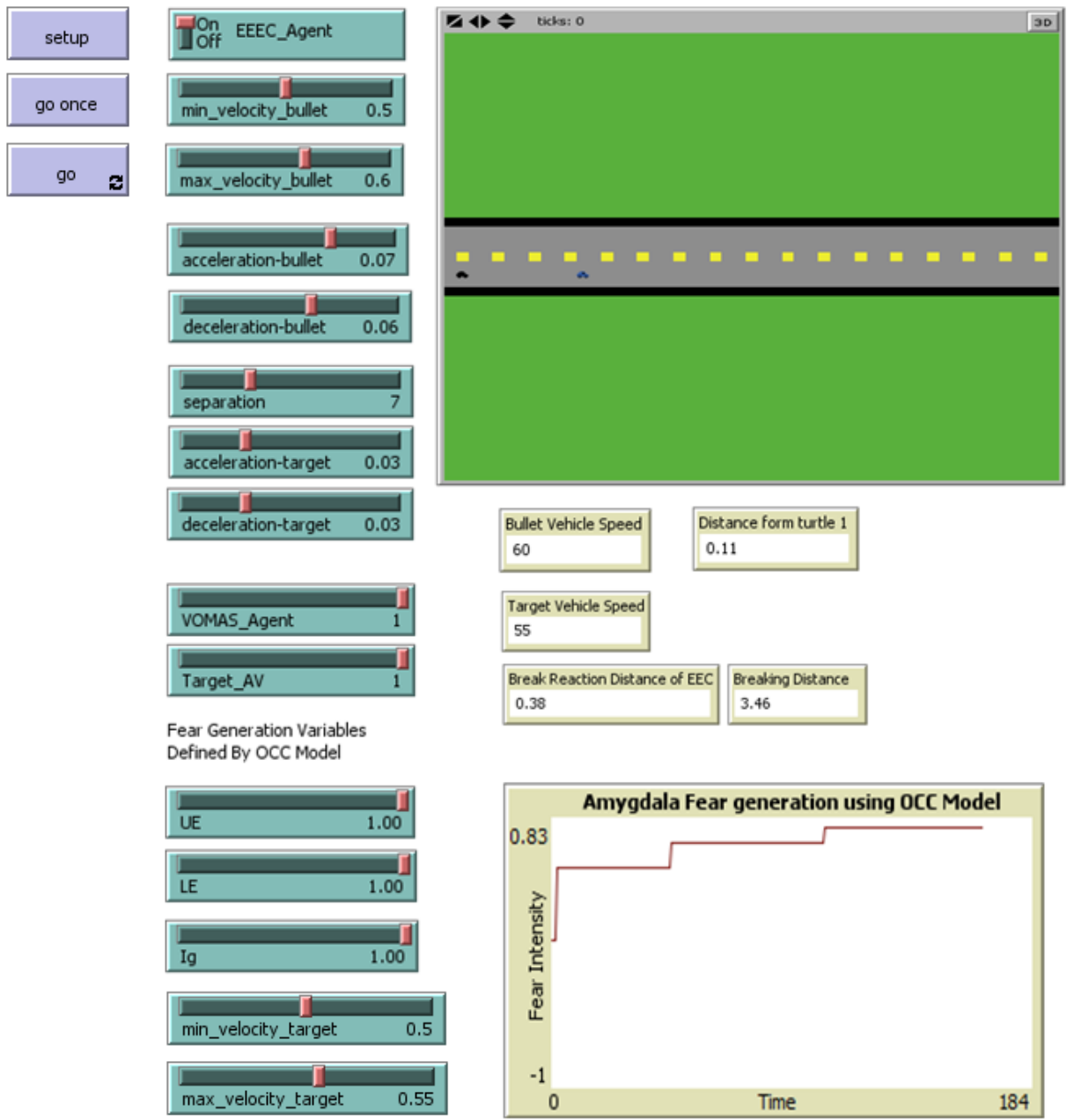

Figure. 6 Main Simulation Screen of EEEC_Agent based Collision Avoidance system in NetLogo Environment

### a) Experiments-Invariant1_TypeA

In this experiment, we have validated the invariant 1. For this purpose, six different types of tests with different parameter values have been designed as presented in Table 7. These tests are set up to test the pre-condition that if the distance between both AVs will be decreased, what will be its effect on the intensity of the fear of EEEC_Agent. This set of tests has been performed using the behavior space tool of NetLogo 5.3.1 environment and each test has been repeated 50 times.

Table 7 Parameter values for Experiment-Invariant1_TypeA

| Experiment No | Separation | Min Velocity | Max Velocity | Acceleration _ Bullet | Deceleration _Bullet | Acceleration _ Target | Deceleration _Target | EEEC_ Agent |
|---|---|---|---|---|---|---|---|---|
| 1 | 1 | 10 | 100 | 0.06 | 0.03 | 0.03 | 0.03 | True |
| 2 | 1 | 10 | 100 | 0.06 | 0.06 | 0.03 | 0.03 | True |
| 3 | 1 | 60 | 100 | 0.06 | 0.03 | 0.03 | 0.03 | True |
| 4 | 1 | 60 | 100 | 0.06 | 0.06 | 0.03 | 0.03 | True |
| 5 | 1 | 90 | 100 | 0.06 | 0.03 | 0.03 | 0.03 | True |
| 6 | 1 | 90 | 100 | 0.06 | 0.06 | 0.03 | 0.03 | True |

**b). Experiments-Invariant1_TypeB**

To validate the invariant 1_TypeB, a different set of experiments has been defined. This experiment further consists of five tests. These five tests have been designed to check the behavior of Bullet AV (EEEC_Agent) on different distances from the Target AV. The parametric values of 5 tests are presented in Table 8.

Table 8 Parameter values for Experiment-Invariant1_TypeB

| Experiment No | Separation | Min Velocity | Max Velocity | Acceleration _ Bullet | Deceleration _Bullet | Acceleration _ Target | Deceleration _Target | EEEC_Agent |
|---|---|---|---|---|---|---|---|---|
| 1 | 5 | 10 | 100 | 0.06 | 0.03 | 0.03 | 0.03 | True |
| 2 | 9 | 10 | 100 | 0.06 | 0.06 | 0.03 | 0.03 | True |
| 3 | 13 | 10 | 100 | 0.06 | 0.03 | 0.03 | 0.03 | True |
| 4 | 13 | 60 | 100 | 0.06 | 0.06 | 0.03 | 0.03 | True |
| 5 | 17 | 10 | 100 | 0.06 | 0.06 | 0.03 | 0.03 | True |

**c). Experiment-Invariant2**

In this experiment, we have validated the invariant 2. The experiments are set up to test the pre-condition that if the Bullet autonomous vehicle is successfully reacting to the collision threat using EEEC_Agent short route reaction time, then EEEC_Agent requires smaller SSD as compared to the SSD required by Human driver. For this purpose, total 12 experiments have been performed.

**d). Experiment-Invariant3**

In this experiment, we have validated the invariant 3. The experiment is set up to test the pre-condition that if the Bullet autonomous vehicle is successfully reacting to the collision threat using EEEC_Agent short route reaction time, the EEEC_Agent then requires smaller OSD as compared to the OSD required by Human driver. For this purpose, the experiment has been repeated 10 times with different speeds and distances between both AVs.

## 7 Results And Discussion

This section describes the results of both experiments 1 and 2. The results are compared with the state of the art EEC_Agent proposed in [6].

### 7.1 Experiment 1

Table 9 shows the quantitative values of undesirability from very low (VL) to very high (VH). The terms VLD, LD, MD, HD, and VHD represent very low desirability, low desirability, medium desirability, high desirability and very high desirability respectively. If the agent has a value between 0-0.24 for its undesirability of an event, then it can be interpreted as the very low undesirability. However, from an abstract analysis, it can be noted that due to the fuzzy nature of the emotion fear, the boundary of one intensity level mixes in the boundary of another intensity level. Hence, the intensity levels lying between 0.24 and 0.5 will be interpreted as low undesirability and lower than these values as the very low undesirability. In the same way, the other intensity levels of undesirability variable can be interpreted.

In the same way, Table 10 and Table 11 are showing the five quantitative values for finding the different intensity levels of likelihood and Ig variables.

These quantitative values of Desirability, Likelihood and Ig are presented in Table 9, 10 and 11 respectively. These values are then provided to the EEEC_Agent for computing different intensities of fear in the next section by following the proposed SimConnector design.

Table 9 Quantitative Values of Five Intensity levels of Desirable Variable

| VLD | LD | MD | HD | VHD |
|---|---|---|---|---|
| 0-0.24 | 0.1-0.5 | 0.25-0.73 | 0.51-0.9 | 0.76-1 |

Table 10 Quantitative Values of Five Intensity levels of Likelihood Variable

| VLL | LL | ML | HL | VHL |
|---|---|---|---|---|
| 0-0.24 | 0.1-0.5 | 0.25-0.73 | 0.51-0.9 | 0.76-1 |

Table 11 Quantitative Values of Five Intensity levels of Global Variable (Ig)

| VLIg | LIg | MIg | HIg | VIg |
|---|---|---|---|---|
| 0-0.24 | 0.1-0.5 | 0.25-0.73 | 0.51-0.9 | 0.76-1 |

## 7.2 Experiment 2

In this section validation of fear generation mechanism of EEEC_Agent has been performed using VOMAS agent methodology defined in the CABC framework.

### 7.2.1 Results and Discussion: Experiment-Invariant1_TypeA

For the detailed validation of proposed EEEC agent, we have designed two different sets of experiments. Furthermore, these two sets of experiments consist of six and five tests respectively. The details of these tests have been provided in the section 5.1.12. Before starting the discussion, it is important to mention that the simulation for each test within experiments set 1 has been performed for 100 ticks. However, due to the space limitations, the data of only 8-9 ticks have been shown in the Tables 12 to Table 17. However, the graphs shown in figure 7 to figure 12 have been generated over 100 ticks. Furthermore, to draw graphs more clearly, the intensities of fear have been mapped from the range of [0-1] to the [0-100].

The first set of experiments has been designed to validate the performance of EEEC agent on the very short distance between bullet and target AV. The results of the first test of experiments have been presented in Table 12. In the first test, the bullet AV follows the target AV with low speed, i.e. 10 mph with high acceleration rate, i.e. 0.06 mph and low deceleration rate i.e. 0.03 mph.

From the Table 12, it can be seen that at the beginning of travel the distance between bullet and target AV is 2.9 feet and bullet AV requires 0.16 feet as SSD to avoid the collision. Because of small differences between required SSD and current distance from target AV, bullet AV starts feeling high fear i.e. 66. After feeling high fear, the bullet vehicle starts decelerating with the deceleration rate of 0.03 mph and the distance between both AVs starts increasing. From the 2nd entry of Table 12, it can be seen that due to increasing in distance from 2.9 feet to 4.38 feet the bullet AV starts feeling medium fear i.e. 49. From the fifth entry of Table 12 it can be seen that when the required SSD is very near to the distance between both AVs, bullet AV starts exhibiting a high intensity of fear again and as the difference between required SSD and distance moves towards high negative value then the bullet AV starts feeling very high fear. From these tests, it can be validated that the proposed EEEC agent has the capability to feel abrupt fear as well due to the sudden appearance of the leading vehicles on very short distance. The graphical representation of the results of table1 has been shown in figure 7.

Table 12 Computation of Fear at low speed of Bullet AV

| Speed=10 | Acceleration= 0.06 | Deceleration= 0.03 |
|---|---|---|
| **SSD** | **Distance** | **Fear Intensity** |
| 0.16 | 2.9 | 66 |
| 0.35 | 4.38 | 49 |
| 0.76 | 6.36 | 49 |
| 2.61 | 10.36 | 49 |
| 10.23 | 10.89 | 66 |
| 10.23 | 9.74 | 76 |
| 10.23 | 7.0 | 76 |
| 10.23 | 6.0 | 76 |
| 10.23 | 5.50 | 76 |

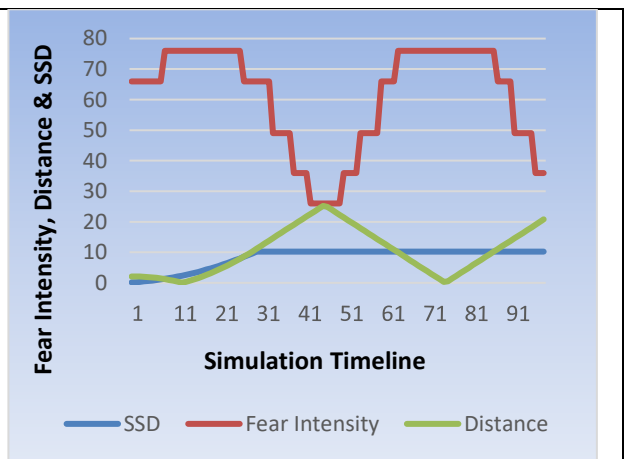

Figure 7 Computation of Fear at low speed of Bullet AV With high acceleration and low deceleration

Table 13 Computation of Fear at low speed of Bullet AV

| Speed=10 | Acceleration= 0.06 | Deceleration= 0.06 |
|---|---|---|
| **SSD** | **Distance** | **Fear Intensity** |
| 0.16 | 2.9 | 66 |
| 0.16 | 4.47 | 49 |
| 0.16 | 7.61 | 49 |
| 0.16 | 9.95 | 36 |
| 0.16 | 13.09 | 26 |
| 0.16 | 18.6 | 16 |
| 0.16 | 20.17 | 6 |
| 0.16 | 15.93 | 26 |
| 0.16 | 10.50 | 36 |

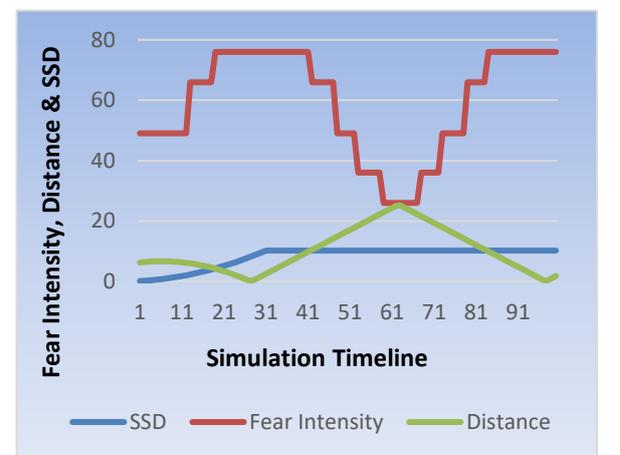

Figure 8 Computation of Fear at low speed of Bullet AV With equally high acceleration and deceleration

Table 14 Computation of Fear at moderate speed of Bullet AV

| Speed=60 | Acceleration= 0.06 | Deceleration= 0.03 |
|---|---|---|
| **SSD** | **Distance** | **Fear Intensity** |
| 3.83 | 2.87 | 76 |
| 5.87 | 3.71 | 76 |
| 9.45 | 3.66 | 76 |
| 10.23 | 3.39 | 76 |
| 10.23 | 2.96 | 76 |
| 10.23 | 2.83 | 76 |
| 10.23 | 2.68 | 76 |
| 10.23 | 2.55 | 76 |
| 10.23 | 2.40 | 76 |
| 10.23 | 2.27 | 76 |

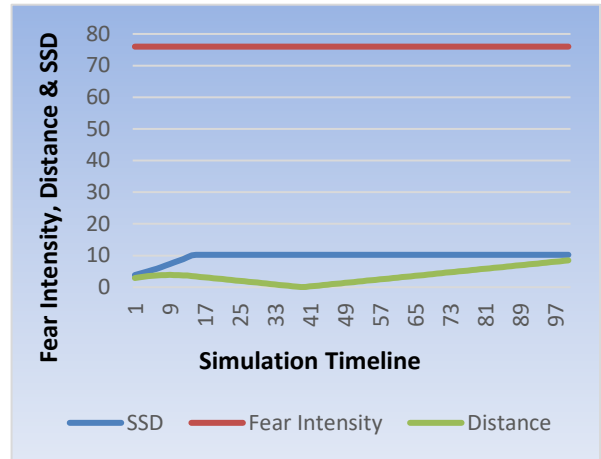

Figure 9 Computation of Fear at moderate speed of Bullet AV
With high acceleration and low deceleration

Table 15 Computation of Fear at moderate speed of Bullet AV

| Speed=60 | Acceleration= 0.06 | Deceleration= 0.06 |
|---|---|---|
| **SSD** | **Distance** | **Fear Intensity** |
| 3.83 | 2.9 | 76 |
| 3.83 | 2.43 | 76 |
| 3.83 | 2.05 | 76 |
| 3.83 | 1.84 | 76 |
| 3.83 | 1.65 | 76 |
| 3.83 | 1.56 | 76 |
| 3.83 | 1.51 | 76 |
| 3.83 | 1.29 | 76 |
| 3.83 | 1.24 | 76 |

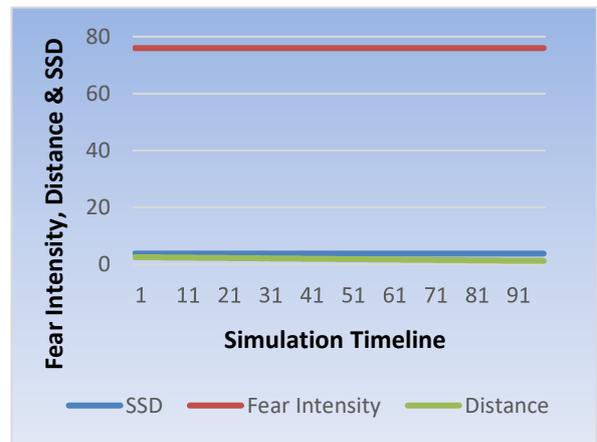

Figure 10 Computation of Fear at moderate speed of Bullet AV with equally high acceleration and deceleration

TABLE 16 Computation of Fear at high speed of Bullet AV

| Speed=90 | Acceleration= 0.06 | Deceleration= 0.03 |
|---|---|---|
| **SSD** | **Distance** | **Fear Intensity** |
| 8.34 | 2.53 | 76 |
| 8.34 | 7.61 | 76 |
| 10.23 | 11.98 | 66 |
| 10.23 | 14.43 | 49 |
| 10.23 | 16.38 | 49 |
| 10.23 | 17.83 | 49 |
| 10.23 | 18.33 | 36 |
| 10.23 | 19.78 | 36 |
| 10.23 | 19.50 | 36 |

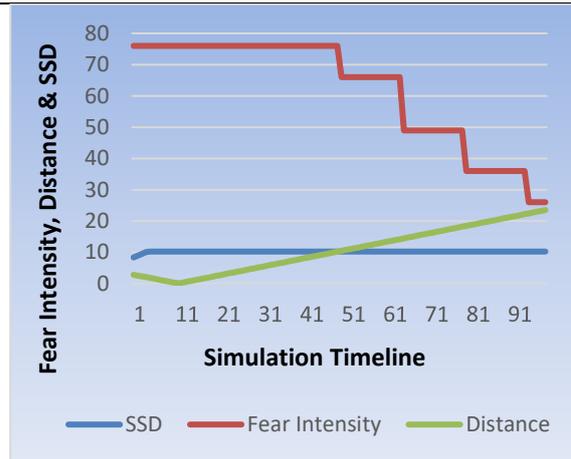

Figure 11 Computation of Fear at high speed of Bullet AV with high acceleration and low deceleration

| TABLE 17 Computation of Fear at high speed of Bullet AV | | |
|---|---|---|
| Speed=90 | Acceleration= 0.06 | Deceleration= 0.06 |
| SSD | Distance | Fear Intensity |
| 8.34 | 2.48 | 76 |
| 8.34 | 7.12 | 76 |
| 8.34 | 10.18 | 66 |
| 8.34 | 12.8 | 49 |
| 8.34 | 16.71 | 36 |
| 8.34 | 20.65 | 26 |
| 8.34 | 20.51 | 26 |
| 8.34 | 20.76 | 26 |
| 8.34 | 20.50 | 26 |

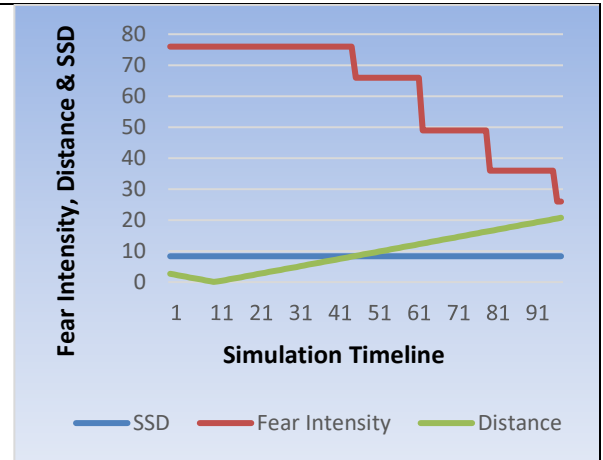

Figure 12 Computation of Fear at high speed of Bullet AV with equally high acceleration and deceleration

Table 13 presents the simulation results of test 2. In test 2, the initial distance between the bullet-target AVs and the speed of bullet AV has been considered same as test 1. However, the acceleration and deceleration rates of bullet AV have been set to high rate i.e. 0.06 mph. From the Table 13, it can be seen that in the beginning of the simulation, EEEC_Agent enabled bullet AV felt high fear due to small differences between required SSD and current distance. In a result, bullet AV starts decelerating with the rate of 0.06 mph. Due to deceleration maneuver, the distance between both AVs starts increasing. In the result, the EEEC agent's fear state switches from high fear to medium fear i.e. 49. An interesting thing in this test can be noted that the SSD remains constant. It is because of equal high acceleration and deceleration rates. From the 6th entry of Table 13, it can be observed that when the distance between both AVs gets very high i.e. 18.6 feet (on simulation scale) as compared to the required SSD i.e. 0.16 feet then the EEC agent fears state presents very low fear i.e. 16. Again, the results of test 2 validate the performance of proposed EEEC agent over very short, medium and high distances between bullet and target AVs. The graphical representation of the results of table 13 has been shown in figure 8.

Table 14 presents the simulation results of test3, which is designed to test the behavior of the EEEC_Agent over moderate speed, i.e. 60 mph with acceleration rate, i.e. 0.06 mph and low deceleration rate i.e. 0.03 mph. From the first entry of table 14, it can be seen that over moderate speed, bullet AV starts feeling very high fear at the very beginning of travel due to small differences between required SSD and current distance. In the next move, bullet AV starts decelerating to avoid the collisions and the distance between both vehicles starts increasing. But still, the EEEC_Agent remains in the very high fear state because of the high danger of accident as compared to the test 1 and test 2 results, where over low-speed EEEC_Agent switches from high fear state to medium fear state after performing deceleration maneuver. The results of test 3 and test 4 are almost same. However, the main difference is that in the test 4 the required SSD remains same for each tick of simulation due to equally high acceleration and deceleration rates. The graphical representations of table 14 and table 15 are presented in figure 9 and figure 10 respectively.

Table 16 and 17 present the results of test 5 and 6 respectively. Test 5 and 6 are designed to validate the performance of proposed EEEC_Agent over very high speed i.e. 90 mph with high acceleration and low deceleration rates and with equal high acceleration and deceleration rates respectively. From the table 16, it can be seen that the proposed EEEC_Agent switches between different fear levels, according to the low or high differences between required SSD and current distance. The graphical representations of table 16 and 17 are presented in figure 11 and figure 12 respectively. From all of these tests it can be concluded that the proposed EEEC_Agent has been validated to have the capability to feel sudden fear over the different speeds with small initial SSDs. Hence, these results validate the invariant1_Type A claim that if the pre-condition that **"*Distance between the rear end of the first AV and the front end of the second AV is very small*"** is true, then the fear level exhibited by AV would result in a post-condition of **"*Intensity of a fear of a bullet autonomous vehicle is high or very high*"**.

### 7.2.2 Results and Discussion: Experiment-Invariant1_TypeB

Further validation of the proposed EEEC_Agent has been provided through an extensive set of tests over different arrangements of experiments. Continuing the discussion section, the rest of the validation scenarios have been presented through table 18-22. Five different sets of experiments have been designed over different initial separation, covering a range from short to long distances between bullet and target AVs. In Table 18, results have been given for validating the EEEC_Agent by placing bullet and target AVs 5 separation apart. Bullet AV is moving at a low speed of 10 mph and accelerating at a rate of 0.06 mph and decelerating with a low rate of 0.03 mph. At tick number 1, which is shown by the first record in table 18, bullet AV requires 0.16 feet SSD. Whereas, the actual distance, i.e. 6.28 feet between vehicles is greater than the required SSD. That is why medium level fear is felt by EEEC_Agent i.e. 49. As the bullet AV proceeds further by adding 0.06 mph to its current speed, a decrease in the distance has been recorded, which is shown by the second entry of distance in table 18 i.e. 5.77 feet. This decrease in distance increased fear intensity and shifted it to the high level. As bullet AV continues to accelerate the require SSD varies with changes in speed on every tick. The fourth record in table 18 is showing the status of bullet AV with an increased SSD value which is 4.73 due to increasing in its speed. At this point, the autos' separation has crossed the safety sight distance limit, which is causing our EEEC_Agent to feel high positive fear i.e. 76. After the violation of SSD, bullet AV tends to decelerate. The rest of the entries are confirming the fact that deceleration causing an increment in distance and an ultimately decrement in fear level. Graphical representation of all 100 ticks' data has been provided in figure 13.

| Table 18 Computation of Fear at low speed of Bullet AV with separation 5 | | |
|---|---|---|
| Speed=10 | Acceleration= 0.06 | Deceleration= 0.03 |
| **SSD** | **Distance** | **Fear Intensity** |
| 0.16 | 6.28 | 49 |
| 2.61 | 5.77 | 66 |
| 3.95 | 4.60 | 66 |
| 4.73 | 3.83 | 76 |
| 10.22 | 7.61 | 76 |
| 10.22 | 11.98 | 66 |
| 10.22 | 14.15 | 49 |
| 10.22 | 18.15 | 36 |
| 10.22 | 17.50 | 36 |

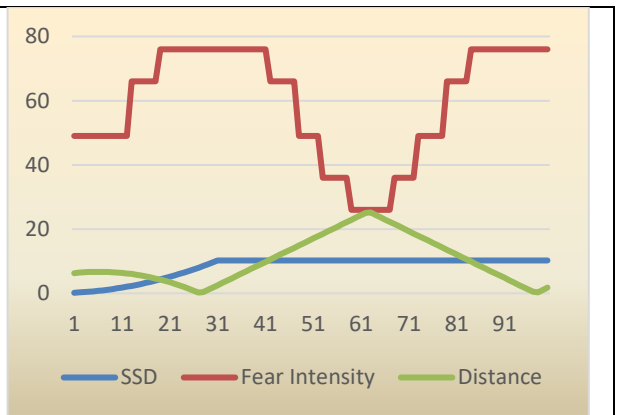

Figure 13 Computation of Fear at low speed of Bullet AV With high acceleration and low deceleration

| Table 19 Computation of Fear at low speed of Bullet AV with separation 9 | | |
|---|---|---|
| Speed=10 | Acceleration= 0.06 | Deceleration= 0.06 |
| **SSD** | **Distance** | **Fear Intensity** |
| 0.16 | 10.24 | 36 |
| 0.16 | 12.21 | 26 |
| 0.16 | 13.27 | 26 |
| 0.16 | 16.44 | 16 |
| 0.16 | 19.49 | 16 |
| 0.16 | 19.60 | 16 |
| 0.16 | 20.14 | 160.15 |
| 0.16 | 21.87 | 6 |

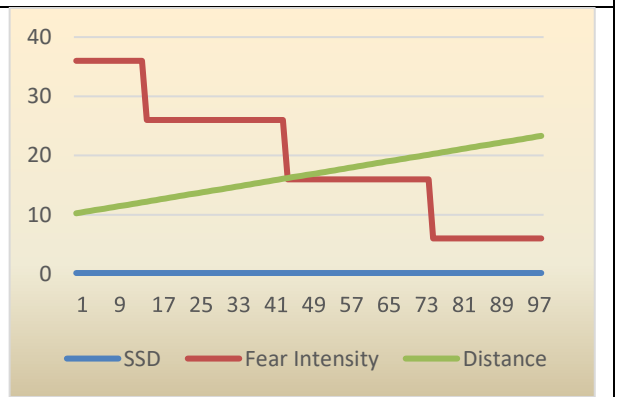

Figure 14 Computation of Fear at low speed of Bullet AV With equally high acceleration and deceleration

| Table 20 Computation of Fear at moderate speed of Bullet AV with separation 13 |||
|---|---|---|
| Speed=10 | Acceleration= 0.06 | Deceleration= 0.03 |
| **SSD** | **Distance** | **Fear Intensity** |
| 0.15 | 14.84 | 26 |
| 0.46 | 16.84 | 16 |
| 5.57 | 23.28 | 16 |
| 7.47 | 23.39 | 26 |
| 10.22 | 21.68 | 36 |
| 10.22 | 18.17 | 49 |
| 10.22 | 14.15 | 66 |
| 10.22 | 11.28 | 66 |

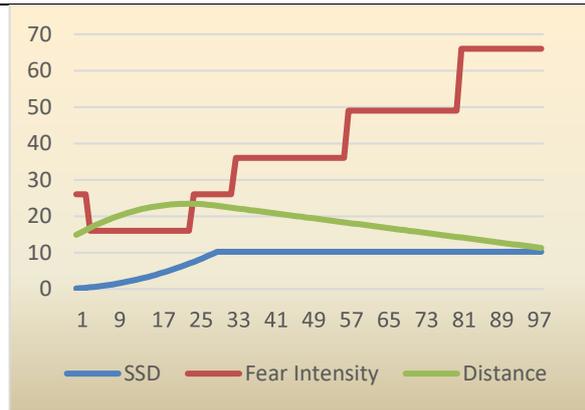

Figure 15 Computation of Fear at moderate speed of Bullet AV With high acceleration and low deceleration

| Table 21 Computation of Fear at moderate speed of Bullet AV with separation 13 |||
|---|---|---|
| Speed=60 | Acceleration= 0.06 | Deceleration= 0.06 |
| **SSD** | **Distance** | **Fear Intensity** |
| 3.83 | 14.9 | 36 |
| 3.83 | 16.04 | 26 |
| 3.83 | 20.03 | 16 |
| 3.83 | 24.02 | 6 |
| 3.83 | 22.99 | 16 |
| 3.83 | 19.30 | 26 |
| 3.83 | 15.31 | 36 |
| 3.89 | 11.59 | 49 |

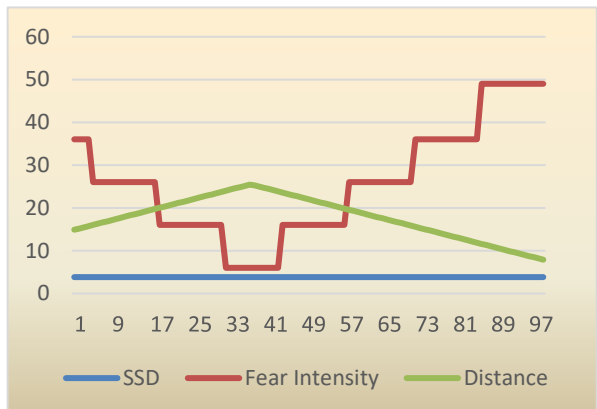

Figure 16 Computation of Fear at moderate speed of Bullet AV with equally high acceleration and deceleration

| Table 22 Computation of Fear at high speed of Bullet AV with separation 17 |||
|---|---|---|
| Speed=10 | Acceleration= 0.06 | Deceleration= 0.06 |
| **SSD** | **Distance** | **Fear Intensity** |
| 0.15 | 18.37 | 16 |
| 0.15 | 20.21 | 6 |
| 0.15 | 22.44 | 6 |
| 0.15 | 24.12 | 6 |
| 0.15 | 25.35 | 6 |
| 0.15 | 20.32 | 6 |
| 0.15 | 19.88 | 16 |
| 0.15 | 18.92 | 16 |

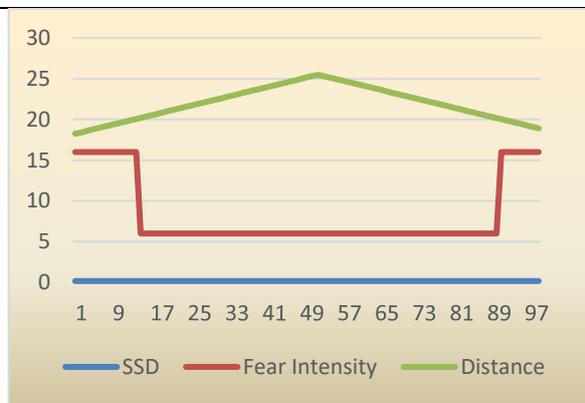

Figure 17 Computation of Fear at high speed of Bullet AV with equally high acceleration and deceleration

Table 19 presents the figures regarding the validation test being performed with a low initial speed of bullet AV, i.e. 10 mph with the initial separation of 9. Bullet AV has equally high acceleration and deceleration rate of 0.06 mph. Initial separation has been increased by 4 points than the previous setup. Staircase representation shown in figure 12 is substantiating the reality of the increase in the distance causes a decrease in the fear intensity. Required SSD for bullet AV is 0.16 feet due to its low speed. Initial placement of the bullet and target vehicles is far apart. The distance shown by the first record in table 19 is 10.24 feet, which has allowed bullet AV to move continuously with mentioned rate by feeling positive low fear given by the value 36. An SSD column of table 19 is baring a constant value due to an equal change in speed caused by both acceleration and deceleration. Deceleration of bullet AV causing an increase in distance, for instance, entry number 3 shows the distance value of 13.27 feet hence lowering the fear value to 26. Fear continues to drop and that is because of the fact that bullet AV is decelerating without any considerable fear intensity. Figure 14 shows a plot of 100 records to show the overall behavior of EEEC_Agent.

In the continuation of the experiments, next test has been performed with the initial separation of 13. The results of this scenario have been shown in table 20 and table 21. Bullet AV is moving with a speed of 10 mph. The first record in table 20 is giving 0.15 feet SSD, 14.85 feet as initial distance and fear value equal to 26 i.e., low fear. Bullet vehicles accelerated, but being on low speed as compared to target AV, an increase in the distance has been recorded given by the second and third record in table 20. Gradual acceleration has put the bullet on a maximum speed of 100 mph and hence SSD required for such a high speed is given by 5th entry i.e. 10.22 feet. The rest of the entries giving a constant figure of 10.22 feet for SSD, but gradual decrease in separation due to the high speed of bullet AV. This continuing reduction in the distance value is causing uplift in the fear intensity, exposing by 6th, 7th and 8th records of table 20. By maintaining the said initial separation another test has been conducted by setting the initial speed of bullet AV to moderate level i.e. 60 mph. Figure 16 depicts the truth of obtaining results that show a constant 3.83 feet SSD value with different distances due to equal acceleration and deceleration rate. By examining the table 21 records, it can be clearly validated that with the increase in distance between vehicles caused (due to the deceleration of bullet vehicle or acceleration of target vehicle) a gradual decline in fear values i.e. 0.36 to 0.26 and then finally to the 6. Record number 5 to 10 reveal the reverse of earlier mentioned fact, as presented clearly in figure 16.

The last subject of discussion regarding experiment tests is describing the results by putting vehicles distant apart, which is shown by the separation of 17 in the simulation. Bullet AV is traveling at a speed of 10 mph with equally high acceleration and deceleration rates i.e. 0.06 mph. The initial low speed of bullet AV is causing an increase in distance of 18.37 feet to 20.21 feet, from 20.21 feet to 25.35 feet and hence lowering the fear from 16 to 6 (shown by record number 1 to 5 in table 22). A gradual decrease in the distance and then a corresponding increase in the fear intensity have been portrayed through the rest of the records of table 22.

Figure 17 shows the actual behavior of all 100 records that were gathered during 100 runs of the simulation. The downward movement of the orange line (fear) with the upward movement of the gray line (distance) is validating the invariant1_TypeB.

### 7.2.3  Results and Discussion: Experiment-Invariant 2

Figure 18 shows the validation results of invariant 2. Total 12 experiments are conducted to validate the claim that If the pre-condition that *"Bullet Agent autonomous vehicle is successfully reacting to the rear end collision threat using EEEC_Agent short route reaction time"* is true, then stopping sight distance require would result in a post-condition of *"EEEC_Agent requires smaller SSD as compared to the SSD required by Human driver"*. The vertical axis of figure 16 presents different SSD values, whereas the corresponding velocities are given on the horizontal axis. From the figure 18, it can be seen that when the bullet AV successfully avoids the collision by traveling at the velocity of 15 mph then it takes SSD= 31.733 feet as a safe distance. Whereas in the same case the human driver needs SSD = 106.015 feet. In the same way, when the bullet AV successfully avoids the collision by traveling at a high velocity of 50 mph then it takes SSD= 277.184 feet as a safety distance. Whereas for the same case the human driver needs SSD = 524.790 feet. Hence, the results prove that the claim of invariant 2 is true and the emotions inspired collision avoidance module is working properly in case of rear-end collisions.

The EEC_Agent proposed in [6] is tested for lateral collisions at different speeds. However, the [6] failed to validate the results of the simulation against any standard collision avoidance parameters. We have validated our claim of rear end collisions against the standard Stopping Sight Distance by adapting VOMAS agent approach defined in CABC framework.

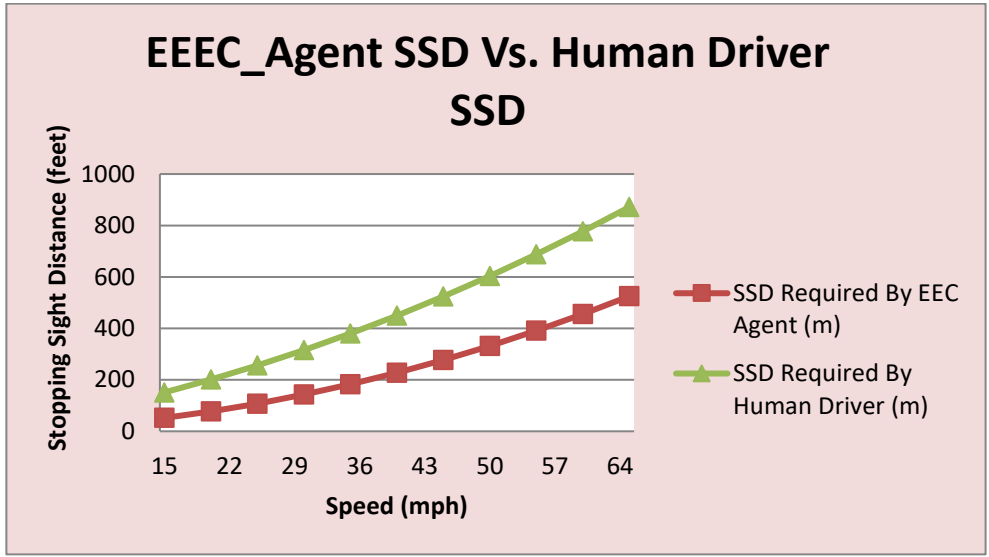

Figure 18 EEEC_Agent SSD vs. Human Driver SSD Graph

### 7.2.4 Results and Discussion: Experiment-Invariant 3

Figure 19 shows the validation results of invariant 3. Total 7 experiments are conducted to validate the claim that if the pre-condition that *"Bullet vehicle is successfully reacting to the overtaking collision threat using EEEC_Agent short route reaction time"* is true, then overtaking sight distance required would result in a post-condition of *"EEEC_Agent requires smaller OSD as compared to the OSD of the human driver"*. From the figure 19 it can be seen that when the bullet AV successfully avoids the collision by traveling at a velocity of 25 mph then it takes OSD= 63.408 feet as a safe distance. However, for the same case, the human driver needs OSD = 85 feet. In the same way, when the bullet AV successfully avoids the collision by traveling at a higher velocity of 50 mph then it takes OSD= 145.264 feet as a safe distance. Whereas for the same case the human driver needs OSD = 185 feet. Hence, the results prove that the claim of invariant 3 is true and the emotions inspired collision avoidance module is working properly in case of overtaking.

The EEC_Agent proposed in [6] is tested for lateral collisions at different speeds. However, the [6] failed to validate the results of the simulation against any standard collision avoidance parameters. We have validated our claim of overtaking collisions against the standard Overtaking Sight Distance parameters by adapting VOMAS agent approach defined in CABC framework.

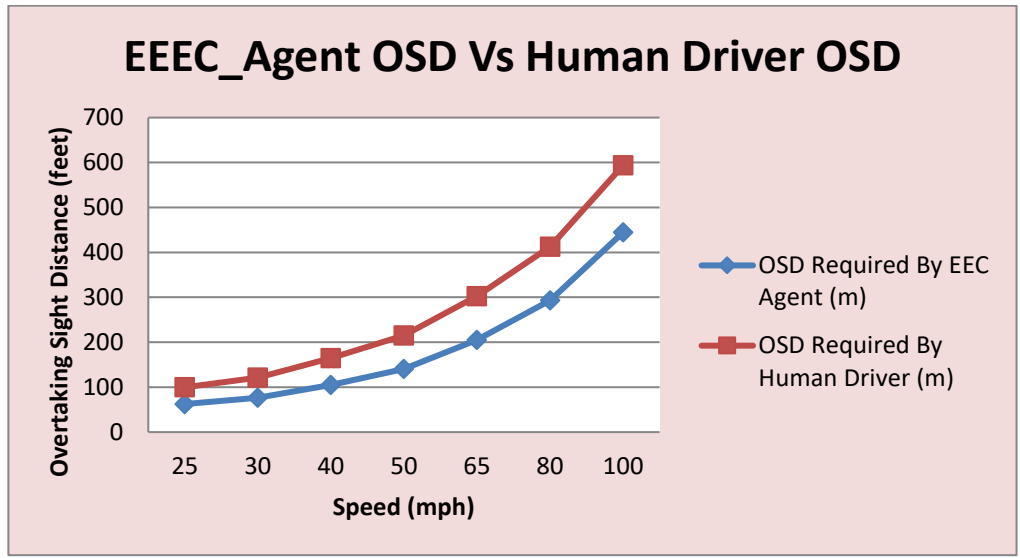

Figure 19 EEEC_Agent OSD vs. Human Driver OSD Graph

## 8 Practical validation of the Validation of Fear Generation Mechanism of EEEC_Agent

To further validate the results of VOMAS agent based simulation regarding fear generation, EEEC_Agent agent has been deployed in a prototype autonomous vehicle. OCC model based application has been developed in the Visual C# environment and it has been deployed in AV using Windows 10 based tablet. Different AV-Obstacle tests have been performed to validate the results of VOMAS agent based simulation regarding the fear generation mechanism of EEEC_Agent. The overall architecture of rigorous validation using VOMAS agent has been presented in figure 20. The prototype AV is equipped with high range ultrasonic sonars, which help to measure the distance of AV from incoming obstacle. To control these sonars, Arduino ATmega2560 processor has been utilized, which further pass the results of these sonars to the C# application, which computes the intensity of fear. To validate that the fear generation processes of EEEC_Agent is practically generating the same results as it generated in NetLogo simulation, VOMAS agent has been considered, which further utilized the invariants defined in section 5.1.

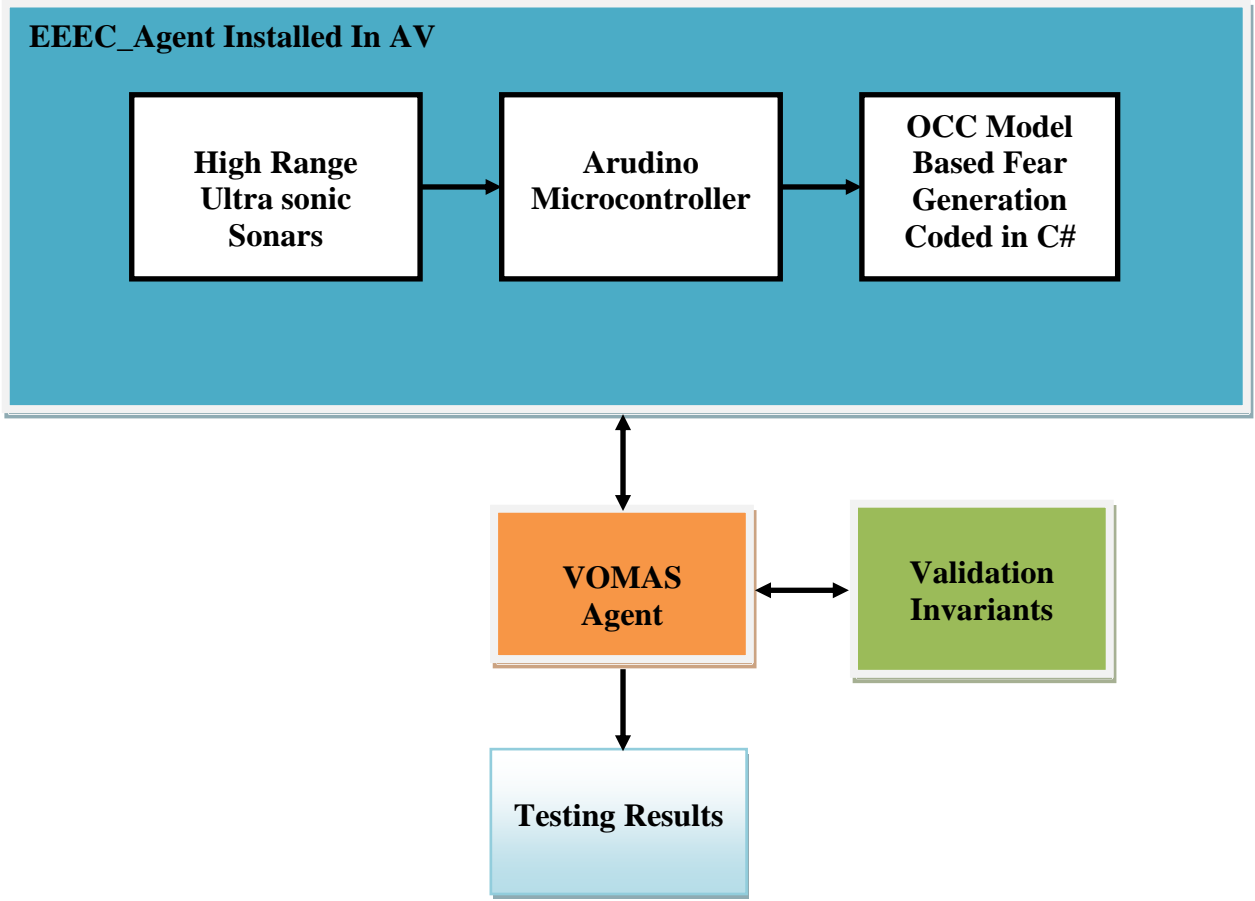

Figure 20 Proposed Practical Validation Architecture Using VOMAS Agent Approach

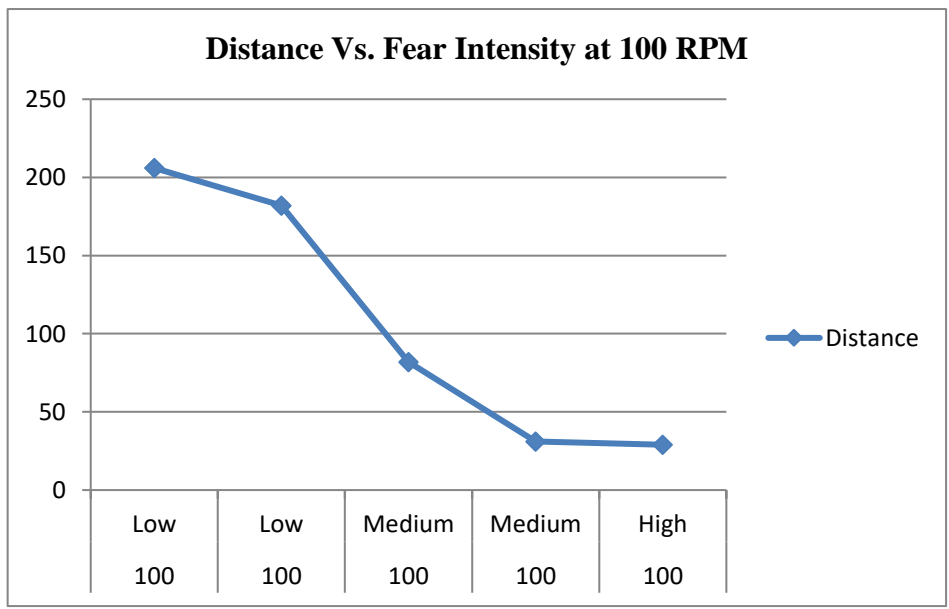

(a)

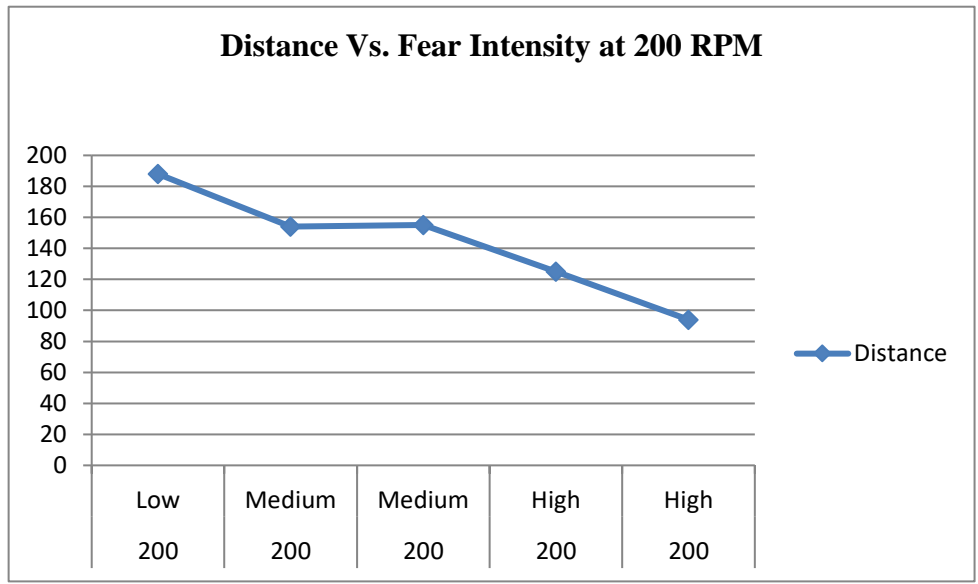

(b)

Figure 21 graphic representations of practical validation results a. Different Fear Intensity levels perceived by EEEC_Agent installed AV at 100 RPM speed, b. Different Fear Intensity levels perceived by EEEC_Agent installed AV at 200 RPM speed

Figure 21 a and b are presenting the results of different practical validation tests, which have been performed using AV-Obstacle topology. From the figure 21a, it can be seen that when the AV moves at low speed towards an obstacle, then in the beginning, it feels low fear due to high distance i.e. 210 feet. Then gradually fear increases as the distance decreases. This confirms the simulation results that the fear generation mechanism of EEEC_Agent is working properly. Further, if we analyzed the results of figure 21b then it can be seen that at high speed EEEC_Agent first feel low fear and then suddenly jump to the medium fear because the distance suddenly decreased from 190 feet to 155 feet. In the same way, other results confirm that the fear increases as the distance between EEEC_Agent and obstacles decreases. An interesting phenomenon can be noted from the figure 21a and 21 b that at low speed, the EEEC_Agent feel medium fear after some time, whereas with high speed the EEEC_Agent suddenly perceive medium fear after feeling low fear. It shows the validation of the proposed fear generation mechanism that fear feeling capability is not fixed but dynamic. Hence, it can be seen that the validation of validation has proved that the fear generation mechanism of proposed EEEC_Agent is working accurately according to the different situations. These results can be further utilized by the researchers to build the EEEC_Agent based systems with more confidence.

# 9 Conclusion

A validated Enhanced Emotion Enabled Cognitive Agent (EEEC_Agent) has been proposed to avoid the collision between autonomous vehicles. For this purpose, prospect based emotions defined by OCC model are used to generate fear in EEEC_Agent. SimConnector approach is used to join fuzzy logic environment results with NetLogo agent-based simulation. VOMAS Agent is used to validate the performance of EEEC_Agent in the rear end and overtaking collisions. The extensive experiments proved that validated EEEC_Agent can perform collision avoidance with smaller SSD and OSD as compared to the human driver.

**Conflict of Interest statement**

The authors declare that there is no conflict of interest in this manuscript.